\documentclass[12pt,preprint]{aastex}
\usepackage{psfig,natbib}
\begin{document}

\newcommand\degd{\ifmmode^{\circ}\!\!\!.\,\else$^{\circ}\!\!\!.\,$\fi}
\newcommand{\etal}{{\it et al.\ }}
\newcommand{\uv}{(u,v)}
\newcommand{\rdm}{{\rm\ rad\ m^{-2}}}

\title{A Radio Survey for Linear and Circular Polarization in Low Luminosity Active Galactic Nuclei}

\author{Geoffrey C. Bower\altaffilmark{1,2}, 
Heino Falcke\altaffilmark{3} \&
Richard R. Mellon\altaffilmark{4}}

\altaffiltext{1}{National Radio Astronomy Observatory, P.O. Box O, 1003 
Lopezville, Socorro, NM 87801; gbower@nrao.edu} 
\altaffiltext{2}{Radio Astronomy Laboratory, 
University of California, Berkeley, CA 94720; gbower@astro.berkeley.edu}
\altaffiltext{3}{Max Planck Institut f\"{u}r Radioastronomie, Auf dem 
H\"{u}gel 69, D 53121 Bonn Germany; hfalcke@mpifr-bonn.mpg.de} 
\altaffiltext{4}{Department of Astronomy and Astrophysics, 
Pennsylvania State University, University Park, PA 16803; rmellon@astro.psu.edu}

\begin{abstract}

We conducted
a Very Large Array survey of eleven low luminosity active galactic
nuclei for linear and circular polarization at 8.4 GHz.  We detected circular
polarization in one source (M81*) and linear polarization in 3 sources.
Sensitivity limits were $\sim 0.1\%$ for both modes of polarization in 9 of 11
sources. The detections confirm the importance of nonthermal emission in
LLAGN.  However, detection rates for circular and linear polarization are lower for these
sources than for more powerful AGN.  Fractional linear polarization in detected sources is also 
lower than in more powerful AGN.
The weak linear polarization in the survey sources indicates their overall similarity to Sgr A*.
Confusion with thermal sources, depolarization and weaker, less extended
jets may contribute to these differences.  We detect a rotation measure $\ga 7 \times 10^4 \rdm$
for NGC 4579.  This may arise from magnetized plasma in the accretion, outflow or interstellar
regions.  Inverted spectra 
are present in both M81* and Sagittarius A* and absent from all sources in which circular polarization
is not detected.  This suggests that optical depth effects are important in the creation of
circular polarization.  

\end{abstract}


\section{Introduction}

Circular and linear polarization are important diagnostics of nonthermal radio emission
and its environment in
extragalactic radio sources and galactic micro-quasars.  In active
galactic nuclei (AGN), linear polarization (LP) has been used to demonstrate
that the emission mechanism is synchrotron radiation, that 
magnetic fields are present and that shocks propagate in jets, compressing
the magnetic fields \citep[e.g.,][]{1985ApJ...298..114M}.   LP is typically greater than 1\%
and is present in a very high fraction of powerful extragalactic radio
sources \citep{1992ApJ...399...16A}.  Rotation measures are used to study
the plasma density and magnetic field in
accretion and outflow regions of AGN \citep{2000ApJ...533...95T}.

The significance of circular polarization (CP) is less clear.  A variety
of mechanisms can produce the small levels of CP
seen in these sources, including intrinsic and extrinsic origins
\citep{1977ApJ...214..522J,2000ApJ...545..798M,2000AAS...197.8315B,
2002A&A...388.1106B,2002ApJ...573..485R}.   
Synchrotron and
cyclotron emission produce CP in varying degrees depending on the the electron
population, magnetic field strength and orientation.  Coherent emission
processes may also lead to CP.  In addition, the 
presence of electrons with Lorentz factors
$\sim 1$ in or near a synchrotron source can significantly alter both
CP and LP through Faraday effects.  The combination of these effects can
lead to complex polarized spectra.
Finally, a magnetized scattering
region can produce scintillating CP.

Recently, there has been a sharp increase in interest in CP observations.
CP has been detected in Sagittarius A* 
\citep{1999ApJ...523L..29B,1999ApJ...526L..85S,2002ApJ...571..843B}, a large number of 
powerful extragalactic radio sources 
\citep{2000MNRAS.319..484R,2001ApJ...556..113H} including some IDV sources 
\citep{2000ApJ...538..623M},
and galactic microquasars
\citep{2000ApJ...530L..29F,2002MNRASinpressfender}.  VLBI imaging
of powerful extragalactic radio sources indicate that the CP is confined
to the cores.  The spectral and variability  properties
in these sources vary widely indicating potentially diverse origins.
In most sources, LP dominates CP, as expected for simple synchrotron
sources.  A notable exception to this is Sgr A*, which exhibits no LP.

The polarization properties of low luminosity AGN (LLAGN) have not
been previously investigated.  
%Rudnick, Jones \& Fiedler (1986) 
\citet{1986AJ.....91.1011R} observed the LP  of 
a sample of ``weak'' radio sources ($L \sim 10^{23} - 10^{26} {\rm\ W\ Hz^{-1}}$)
with flat spectra.  Many of these sources were unpolarized at a level of
1\% at 15 GHz, suggesting that they are not simply scaled down versions of 
powerful AGN.  However, there exists a significant gap in luminosity between
these sources and Sgr A*, which has a radio luminosity $\sim 10^{16} {\rm\ W\ Hz^{-1}}$.

The intrinsic nature of the LLAGN is still in doubt  
\citep[for recent reviews see][]{2001ApJ...562L.133U,2002AApNagarinpress}.  
These sources display emission
line luminosities that may be produced through a variety of mechanisms, including
stellar photoionization, collisional ionization through shocks or in a starburst.
A variety of methods have demonstrated
that some of the sources display AGN-like characteristics.  However, as in the
case of Sgr A*, the luminosities are substantially sub-Eddington and
there is dispute over the roles of advection, convection and outflow in
these sources.  The detection of high brightness temperature radio
components has made a strong case for the role of jets in these sources.
LP and CP can act as probes of the radio-emitting region as well as of 
the accretion and outflow regions through which the emission propagates.

We present here a radio polarization survey of 11 nearby LLAGN with luminosities
in the range $10^{20}$ to $10^{23}$ ${\rm\ W\ Hz^{-1}}$.  In \S 2, we describe the
observations, our error analysis and the results.  CP is only detected
in one source, NGC 3031 (M81*).  We have discussed the significance of this detection
in another paper 
\citep{2001ApJ...560L.123B}.  LP is detected in 4 sources.
We discuss our results in \S 3 and summarize in \S 4.

\section{Observations}

A sample of 11 LLAGN was selected from the samples of
\citet{1991AJ....101..148W} and \citet{1995ApJS...98..477H}.
%Wrobel \& Heeschen (1991) and Ho, Filippenko \& Sargent (1995).  
The selected sources (Table~\ref{tab:results})
are the brightest radio sources in the catalogs
for which we were able to obtain data.

We observed the sample with the Very Large Array in the C
configuration on 4 April 2000 for 12 hours at 8.4 GHz.  The correlator was configured
for continuum observations with 50 MHz bandwidth each in two intermediate frequency bands
in right (RCP) and
left circular polarization (LCP).  The angular resolution was about
$3^{\prime\prime}$.  

Observations were carried out with the same techniques
described in \citet{1999ApJ...523L..29B} and \citet{2002ApJ...571..843B}.
%Bower, Falcke \& Backer (1999) and Bower \etal (2002).
Absolute flux calibration was set with a short observation of 3C 48.
Scans of the target sources were bracketed with scans for 
pointing and calibration on a nearby background compact radio source.  
A second nearby background compact radio source was observed in the
same way as the target sources as a check for CP calibration errors.
We cycled through calibration and target sources in about 10 minutes.
Calibrator, target and check sources were placed in five groups, 
labeled A through E, based on their proximity on the sky.
Integration time for each target source was set to achieve rms noise 
errors of less than 0.1\% except for the weak sources I 1024 and NGC 3516.  
Observations were made over 12 hours,
leading to significant changes in parallactic angle, azimuth angle and
elevation angle for each source.  At least 4 observations were made of
each source.

Data were reduced in AIPS.  The calibration sources were self-calibrated
assuming $V=0$.  Linear polarization calibration was also performed
using the calibration sources.  Amplitude gains, phase gains and polarization
leakage terms were transferred to the target and check sources.  All
sources were imaged in Stokes I, Q, U and V.  Flux densities were estimated 
by fitting Gaussians at the source's location.  Results are summarized
in Table~\ref{tab:results}.  Some sources were resolved in total intensity.
However, polarization images were unresolved for all sources
except NGC 4374 and NGC 4486.  We plot the linear and
circular polarizations together in Figure~\ref{fig:lpcp}.

Systematic errors from beam squint,
gain and leakage terms are estimated according
to the model for the VLA
in \citet{2002ApJ...571..843B} %Bower \etal (2002) 
to be between 0.03 and 0.05\%
in total, depending on the number of scans per source.  We see that 
that measured CP for the check sources range from 0.03 to 0.45\% with
a median of 0.14\%.
The likely source of additional error is polarization of the
calibrator and/or check source.  NGC 3031 is the only source
where the measured CP exceeds that of the check source and is greater than
3 times the thermal noise.  The CP for NGC 3031 is, in fact, 10 times the thermal
noise and repeatedly detectable 
\citep{2001ApJ...560L.123B}.  We do not
count NGC 6500 as a CP detection because of the large measured CP for
the check source, J1745+173.  The two large CP values with the same
sign suggest that the calibrator, J1751+096, may have CP at a level of 
$\sim -0.4\%$.  \citet{1999AJ....118.1942H} report a tentative detection
of $\sim -0.1\%$ for J1751+096 in their 15 GHz VLBA observations, which
is consistent in sign and in magnitude for a steep spectral
index.

\section{Discussion}

\subsection{Confusion}

For all sources, the measured polarization fractions are lower limits
to the polarization of the nuclear source.  For a typical distance of 10 Mpc,
the linear resolution of these observations is $\sim 100$ pc.  
Diffuse thermal radiation can contribute significantly to the total flux.
However, for the sources for which we know of VLBI observations,
the compact radio flux is a substantial component of the total flux.
NGC 3031 and NGC 4486 (M87) both have well-studied VLBI properties, which
show that a compact core dominates the total flux 
\citep{1999Natur.401..891J,2000ApJ...532..895B}.  %Nagar \etal 
\citet{2000ApJ...542..197F} and \citet{2001ApJ...559L..87N} also show that
the ratio of 5 GHz VLBA to VLA flux densities is on the order of unity
for five of our sources: NGC 4278, NGC 4374, NGC 4552, NGC 4579 and NGC 6500.
For all 13 sources in the \citet{2001ApJ...559L..87N} sample, the ratio is greater than
0.48.
Nevertheless, a factor of a few in the ratio of total to compact flux density 
would raise the polarization limits to the point where Sgr A*
and NGC 3031 would not be detected.

\subsection{Linear Polarization}

We detect LP in 3 of 11 sources:  NGC 4278, NGC 4486,
and NGC 4579.  These results show that some LLAGN
do exhibit LP, contrary to the
conclusions of \citet{1986AJ.....91.1011R},
%Rudnick, Jones \& Fiedler 
which were based on less sensitive data.   This is strong evidence for a nonthermal 
contribution to the flux density.   

However, the rate of detection of LP is lower than that of more
powerful AGN.  For example, \citet{2001ApJ...556..113H} detect LP in all but 3
of their 40 sources with a mean value of a few percent.  In addition, the fraction of LP
detected in our sources is lower than in more powerful AGN.  
Excluding M87, the mean fractional LP is $\sim 0.2\%$.  Of the 62
sources in the Pearson-Readhead flux-limited sample,  39 have polarization
greater than 1\% at 4.8 GHz and 57 have polarization greater than 1\%
at 15 GHz \citep{1992ApJ...399...16A}.  Mean polarizations are on the order of a few percent.
These trends are true for quasars, BL Lac objects and
radio galaxies in the sample.  This difference cannot be explained by
confusion.  All three sources have been previously detected with
VLBI and been shown to be dominated by their compact flux
\citep{1999Natur.401..891J,2001ApJ...559L..87N}.

Are the LLAGN intrinsically weakly polarized or are they depolarized
by their surrounding medium?  A number of scenarios may lead to intrinsic depolarization.
Extreme magnetic field disorder in the source will leave the source weakly polarized.
This may be due to the absence of shocks in jets that order the field and produce regions
of high polarization fraction.  A high degree of symmetry in the magnetic field, i.e., 
a face on azimuthal field or a radial field will also lead to a weakly polarized
source.  Such field configurations might be typical of disk accretion or
quasi-spherical accretion.  Finally, internal Faraday depolarization may occur 
if the density of nonrelativistic
particles in the emitting region is large enough 
\citep[e.g.,][]{2002A&A...388.1106B,2002ApJ...573..485R}.

On the other hand, 
depolarization in the accretion, outflow or surrounding regions
is more than adequate to account for this result.
Beam and bandwidth depolarization at these frequencies and in these bandwidths
will occur for rotation measures of
$10^3 \rdm$ and $10^5 \rdm$, respectively.  \citet{2000ApJ...533...95T} has 
measured RMs as large as $2\times 10^3 \rdm$ towards the cores of powerful AGN.
\citet{2002ApJBowersubmitted} have measured a RM$\sim 4\times 10^5 \rdm$
towards Sagittarius A*, which may arise from the accretion and outflow regions
or from an interstellar halo.  Even for the very low accretion
rates of ADAF and CDAF models of Sgr A*, RMs can easily reach $10^6 \rdm$
and higher \citep{1999ApJ...521..582B,2000ApJ...545..842Q}.  
The radio emission from our sources almost certainly
propagates through similar media.  In the case of NGC 4258, for example, the radio
emission originates from within 0.02 pc of the black hole 
\citep{1998ApJ...497L..69H}.  On the other hand, the radio emission from much
more powerful AGN may begin parsecs away from the black hole.

The radio intermediate quasar III Zw 2 provides an interesting case for 
comparison.  This compact radio source ($<1 $ pc) shows superluminal
expansion and spectral evolution
consistent with shock excitation in a relativistic outflow \citep{2001ApJ...560L.123B}.  
However, the source exhibits no linear polarization in 20 years of
monitoring (M. Aller, private communication).

In order  to look for the effects of bandwidth depolarization and large RMs,
we compared the difference between lower and upper sideband position
angles for all sources.  These sidebands are each 50 MHz wide and separated from
each other by 50 MHz.  We find a significant change only
for NGC 4579 of $60 \pm 13$ degrees.  Systematic error due to leakage term
miscalibration is less than 10 degrees.
By comparison,  NGC 4278, which is polarized at a level similar to
NGC 4579, shows a difference of 6 degrees in the position angle.
Changes in the position angle for calibrator sources were also consistent with
thermal noise.
The implied RM for NGC 4579 is $\sim 7 \times 10^4 \rdm$.  This would lead to
bandwidth depolarization $\sim 0.5$, which does not significantly increase
the intrinsic LP.  However, one cannot rule out a much higher RM inducing
more substantial depolarization as well as the angular rotation.  We can
compute an upper limit for the RM $\sim 10^7 \rdm$ with the requirement of
a maximum of 70\% intrinsic polarization.

As discussed
above, this RM could arise from a variety of sources.  Without more detailed knowledge
of the source geometry, it is difficult to place limits on specific
source models such as jet, ADAF and CDAF models.  Unlike the case of millimeter wavelength
polarization in Sgr A*, the centimeter wavelength emission from NGC 4579
may originate far outside of the accretion region.    However, if the emission does
pass through the accretion region, then the accretion rate must be very low
\citep{1999ApJ...521..582B,2000ApJ...545..842Q}.

Spectropolarimetry, higher frequency polarimetry and VLBI
polarimetry are potentially useful tests of these scenarios.  They
could provide quantitative information about field strengths
and particle densities in the inner few parsecs of these sources.

\subsection{Circular Polarization}

We compare the rates of detection of CP here with that in powerful
extragalactic radio sources.  At a level of $\sim 0.1\%$, we detect
1 out of 9 sources.  \citet{2000MNRAS.319..484R} detect CP greater than 0.1\%
in 12 out of 30 sources.  \citet{2001ApJ...556..113H} detect CP greater than 0.1\%
in 11 out of 40 sources.  Due to the small number of sources in our
sample, we cannot conclude that the rate of CP in LLAGN is less than
that in powerful AGN.  However, the trend appears to be that way.

What physical condition leads to the presence of CP in some sources
and not in others?  NGC 3031 and Sgr A* are both unique in that they have inverted
spectra at the frequencies where CP is detected.  We summarize the 
spectral index information for sources where simultaneous spectra are
available in Figure~\ref{fig:lpcp}.  All spectral indices come from
the measurements of \citet{2001ApJ...559L..87N}.  While the number of sources is small,
the trend is apparent.

This effect may be the 
result of an inhomogeneous synchrotron source in the optically
thick regime \citep{2002A&A...388.1106B,2002ApJ...573..485R}.  
Orientation effects may also be of importance since the
magnitude of CP from a synchrotron source depends on the line of
sight magnetic field strength.  Finally, NGC 3031 is also the closest of our
survey sources, suggesting that resolution of thermal sources in the
nucleus and of extended optically, thin regions in the jet may
be important.

\section{Conclusions}

We have observed linear and circular polarization in a sample of low luminosity
AGN.  The polarization properties of these sources differ from those
of more powerful AGN.  Linear polarization is less frequently detected
and less strong when detected.  Circular polarization is less frequently
detected.  The conclusion is not as simple as claiming that these sources
differ intrinsically from the more powerful AGN, however.  While increased
field disorder in the sources could explain the results, environmental
depolarization effects could also play a role.

These polarization measurements probe regions on a scale of parsecs and smaller
in these sources.  Detection of rotation measures in these sources is a direct
measure of the environment of a supermassive black hole, which may be used
to constrain models for accretion and outflow.

A broader sample of sources is necessary to demonstrate
that the trends we see are statistically significant.  VLBI
polarimetric imaging and spectropolarimetry may also test some
of these conclusions.

\acknowledgements The National Radio Astronomy Observatory is a facility of the 
National Science Foundation operated under cooperative agreement 
by Associated Universities, Inc.
This research has made use of data from the University of Michigan Radio Astronomy 
Observatory which is supported by funds from the University of Michigan.

%\bibliographystyle{apj}
%\bibliography{../../myrefs}

\plotone{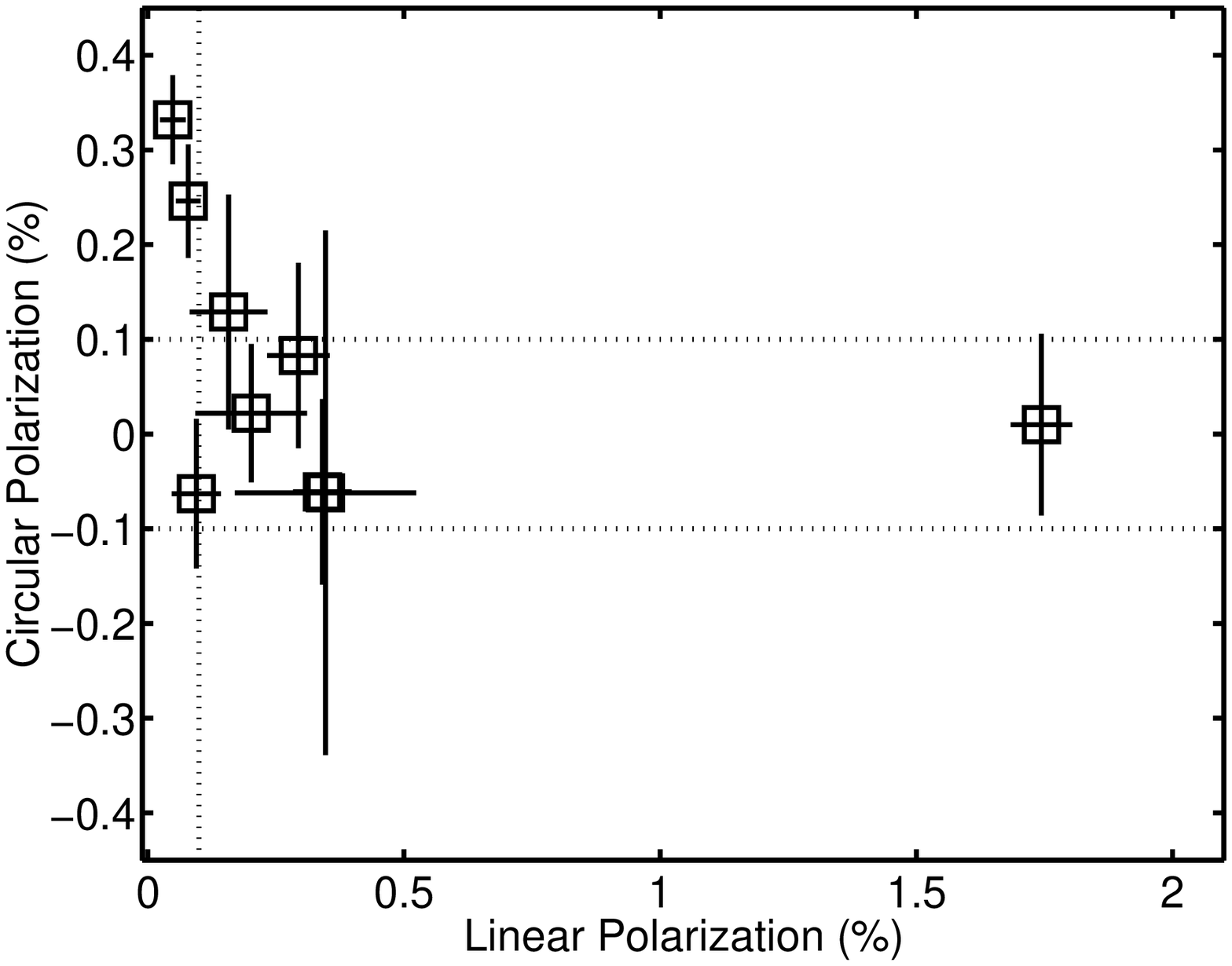}
\figcaption[f1.eps]{Linear and circular polarization for the survey
sources.  Plotted error bars are $1-\sigma$ thermal errors.  Dotted lines
indicate the characteristic systematic errors in linear and circular polarization
of 0.1\%.  We have excluded non-detections for I 1024 and NGC 3516 due to the
substantially larger fractional errors for these sources.  \label{fig:lpcp}}

\plotone{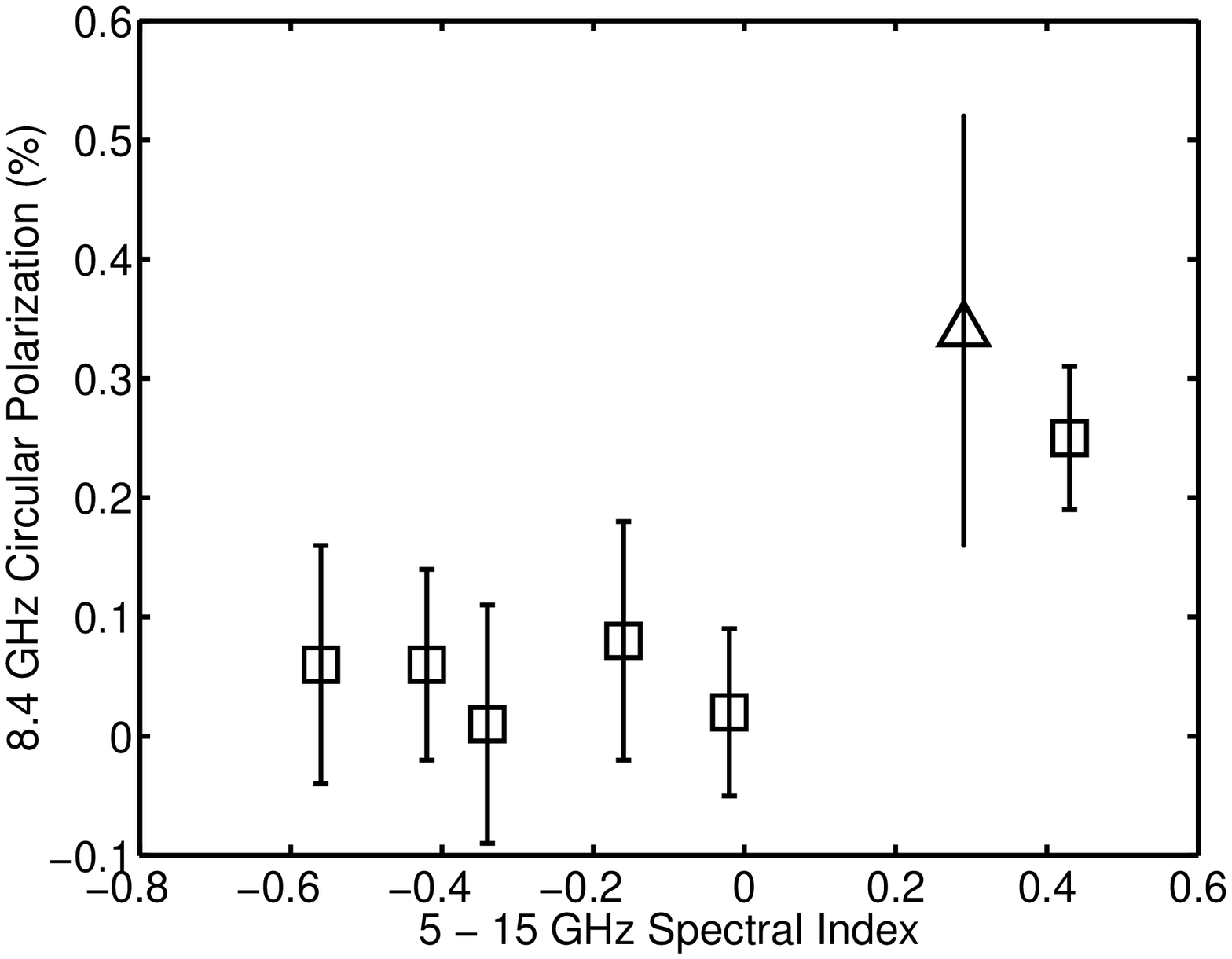}
\figcaption[f2.eps]{Absolute value of circular polarization at 8.4 GHz as a function of spectral
index between 5 and 15 GHz.  
Squares are for sources in the sample for which 5 to 15 GHz contemporaneous
flux density measurements are available.  The triangle is for Sgr A*.
\label{fig:spindex}}

\begin{deluxetable}{cllrrrrr}
\tabletypesize{\footnotesize}
\tablecaption{Polarized and Total Flux from VLA Continuum Observations 
at 8.4 GHz\label{tab:results}}
\tablehead{
\colhead{Group} & \colhead{Source}  & \colhead {Source Type} & \colhead{$I \pm \sigma$} & \colhead{$V \pm \sigma$} & 
\colhead{$P \pm \sigma$} & \colhead{$\chi \pm \sigma$} & \colhead{Resolved} \\
     &            &             & \colhead{(mJy)} & \colhead{(\%)} & \colhead{(\%)} & \colhead{deg} &\colhead{(Y/N)} \\
}
\startdata

 A & 1048+717 & calibrator   & $ 1541.6 \pm    0.2$ & $ -0.00 \pm  0.01$ & $  0.89 \pm  0.00$ & $  -21.0 \pm    0.2$ & N \\ 
 A & 1056+701 & check source & $  467.5 \pm    0.6$ & $  0.14 \pm  0.03$ & $  1.76 \pm  0.01$ & $   81.5 \pm    0.3$ & N \\ 
 A &    NGC 3031 & SrBI-II      & $  263.1 \pm    0.5$ & ${\bf  0.25 \pm  0.06}$ & $  0.08 \pm  0.02$ & $  -29.8 \pm    6.1$ & N \\ 
 A &    NGC 3516 & RSBO$_2$     & $    2.9 \pm    0.0$ & $ -0.76 \pm  1.25$ & $  0.83 \pm  0.59$ & $  -13.3 \pm   23.1$ & Y \\ 
 A &    NGC 4589 & E2           & $   13.0 \pm    0.0$ & $ -0.06 \pm  0.28$ & $  0.35 \pm  0.18$ & $  -10.5 \pm   14.2$ & N \\ 
\\
 B & 1400+621 & calibrator   & $ 1140.8 \pm    0.2$ & $ -0.00 \pm  0.01$ & $  0.15 \pm  0.01$ & $   23.1 \pm    1.1$ & N \\ 
 B & 1335+587 & check source & $  671.5 \pm    1.9$ & $  0.16 \pm  0.04$ & $  0.13 \pm  0.01$ & $   36.3 \pm    1.3$ & N \\ 
 B &    NGC 5216 & E/S0         & $   37.9 \pm    0.1$ & $  0.13 \pm  0.12$ & ${\bf  0.16 \pm  0.08}$ & $   76.0 \pm    6.9$ & N \\ 
\\
 C & 1158+248 & calibrator   & $  555.7 \pm    0.1$ & $  0.01 \pm  0.02$ & $  0.12 \pm  0.01$ & $  -15.0 \pm    2.7$ & N \\ 
 C & 1221+282 & check source & $  732.9 \pm    1.2$ & $  0.03 \pm  0.04$ & $  5.54 \pm  0.02$ & $    5.8 \pm    0.1$ & N \\ 
 C &    NGC 4278 & E1           & $   76.5 \pm    0.3$ & $ -0.06 \pm  0.10$ & ${\bf  0.34 \pm  0.06}$ & $   15.7 \pm    7.0$ & N \\ 
 C & 1239+075 & check source & $  580.4 \pm    1.8$ & $ -0.07 \pm  0.03$ & $  5.26 \pm  0.02$ & $  -56.6 \pm    0.1$ & N \\ 
 C &    NGC 4374 & E1           & $  183.4 \pm    2.4$ & $  0.02 \pm  0.07$ & $  0.20 \pm  0.11$ & $   14.8 \pm   22.3$ & Y \\ 
 C &    NGC 4486 & E0           & $ 3351.3 \pm   30.9$ & $  0.01 \pm  0.10$ & ${\bf  1.74 \pm  0.06}$ & $   22.0 \pm    1.5$ & Y \\ 
 C &    NGC 4552 & SO$_1$       & $   94.1 \pm    0.2$ & $ -0.06 \pm  0.08$ & $  0.09 \pm  0.05$ & $  -39.5 \pm   15.5$ & N \\ 
 C &    NGC 4579 & SabII        & $   45.9 \pm    0.1$ & $  0.08 \pm  0.10$ & ${\bf  0.29 \pm  0.06}$ & $  -32.0 \pm    9.0$ & N \\ 
\\
 D & 1430+107 & calibrator   & $  908.2 \pm    0.2$ & $  0.00 \pm  0.01$ & $  0.04 \pm  0.00$ & $    0.1 \pm    5.1$ & N \\ 
 D & 1445+099 & check source & $  604.9 \pm    0.8$ & $  0.24 \pm  0.03$ & $  1.04 \pm  0.01$ & $   66.6 \pm    0.4$ & N \\ 
 D &    I 1024 & S0           & $    1.4 \pm    0.0$ & $ -1.53 \pm  1.60$ & $  1.96 \pm  1.02$ & $   30.1 \pm   20.9$ & Y \\ 
\\
 E & 1751+096 & calibrator   & $ 2498.2 \pm    0.3$ & $ -0.00 \pm  0.01$ & $  8.33 \pm  0.00$ & $  -25.5 \pm    0.0$ & N \\ 
 E & 1745+173 & check source & $  797.4 \pm    0.9$ & $  0.45 \pm  0.03$ & $  2.87 \pm  0.01$ & $   58.2 \pm    0.1$ & N \\ 
 E &    NGC 6500 & Sbc          & $  138.6 \pm    0.2$ & $  0.33 \pm  0.05$ & $  0.05 \pm  0.03$ & $  -33.2 \pm    9.3$ & N \\ 

\enddata
%\tablecomments{
%Polarized and Total Fluxes from the sources, including the calibrators, check sources, and galaxies.
%The check sources are the quasars used to check the errors on the circular polarization measurements.
%}
\end{deluxetable}

\end{document}